\documentclass[12pt,preprint]{aastex}
\usepackage{epsfig}
\usepackage{amssymb}
\usepackage{graphicx}
\usepackage{natbib}
\usepackage{color}
\bibpunct{(}{)}{;}{a}{}{,}

\newcommand{\Msun}{\mathrm{M}_\odot}
\newcommand{\Rsun}{\mathrm{R}_\odot}
\newcommand{\Mjup}{\mathrm{M}_\mathrm{Jup}}
\newcommand{\Mearth}{\mathrm{M}_\oplus}
\newcommand{\Teff}{\mathrm{T}_\mathrm{eff}}

\defcitealias{kraus_etal2015}{K2015}

\slugcomment{Accepted in ApJ Letters}

\shorttitle{ALMA CO detection and modeling of FW\,Tau} 
\shortauthors{C. Caceres, et al.}

\begin{document}

\title{On the nature of the tertiary companion to FW Tau: ALMA CO observations and SED modeling}

\author{
Claudio Caceres\altaffilmark{1,10},
Adam Hardy\altaffilmark{1,10},
Matthias R. Schreiber\altaffilmark{1,10},
H\'ector C\'anovas\altaffilmark{1,10},
Lucas A. Cieza\altaffilmark{2,10},
Jonathan P. Williams\altaffilmark{3}, 
Antonio Hales\altaffilmark{4,5},
Christophe Pinte\altaffilmark{6,7},
Francois M\'enard\altaffilmark{7,8},
Zahed Wahhaj\altaffilmark{9}
}

\altaffiltext{1}{Instituto de F\'{\i}sica y Astronom\'{\i}a, Universidad de Valpara\'{\i}so, Av. Gran Breta\~na 1111, 2360102 Valpara\'{\i}so, Chile}%
\altaffiltext{2}{N\'ucleo de Astronom\'{\i}a, Universidad Diego Portales, Av. Ej\'ercito 441, Santiago, Chile }%
\altaffiltext{3}{Institute for Astronomy, University of Hawaii at Manoa,  Honolulu, HI 96822, USA}%
\altaffiltext{4}{Atacama Large Millimeter/Submillimeter Array, Joint ALMA Observatory, Alonso de C\'ordova 3107, Vitacura 763-0355, Santiago - Chile}%
\altaffiltext{5}{National Radio Astronomy Observatory, 520 Edgemont Road, Charlottesville, Virginia, 22903-2475, United States}%
\altaffiltext{6}{Univ. Grenoble Alpes, IPAG, F-38000 Grenoble, France CNRS, IPAG, F-38000 Grenoble, France}%
\altaffiltext{7}{UMI-FCA, CNRS/INSU, France (UMI 3386)}%
\altaffiltext{8}{Departamento de Astronom\'{i}a, Universidad de Chile, Camino del Observatorio 1515, Las Condes, Santiago,  Chile}%
\altaffiltext{9}{European Southern Observatory, Av. Alonso de C\'ordova 3107, Vitacura, 19001 Santiago, Chile}%
\altaffiltext{10}{ICM nucleus on protoplanetary disks, Universidad de Valpara\'{\i}so, Av. Gran Breta\~na 1111, 2360102 Valpara\'{\i}so, Chile}

\begin{abstract}
  {\noindent It is thought that planetary mass companions may form
    through gravitational disk instabilities or core accretion.
    Identifying such objects in the process of formation would provide
    the most direct test for the competing formation theories. One of
    the most promising candidates for a planetary mass object still in
    formation is the third object in the FW\,Tau system. We here
    present ALMA cycle 1 observations confirming the recently
    published 1.3\,mm detection of a dust disk around this third
    object and present for the first time a clear detection of a
    single peak $^{12}$CO\,(2--1) line, providing direct evidence for
    the simultaneous existence of a gas disk. We perform radiative
    transfer modeling of the third object in FW\,Tau and find that
    current observations are consistent with either a brown dwarf
      embedded in an edge-on disk or a planet embedded in a low
      inclination disk, which is externally irradiated by the binary
      companion. Further observations with ALMA, aiming for high SNR
      detections of non contaminated gas lines, are required to
      conclusively unveil the nature of the third object in FW\,Tau.}
\end{abstract}


\section{Introduction}

Planetary mass companions (PMC) at wide separations often exceeding
100\,au have been relatively frequently found in recent direct imaging
surveys \citep[e.g.][]{neuhauser_etal2005, lafreniere_etal2008,
  schmidt_etal2008, ireland_etal2011, bailey_etal2014}. If relatively
close, i.e. at separations $\lesssim100$\,au, PMCs may form through
the classical mechanisms of core accretion
\citep{lissauer_stevenson2007} or Class II gravitational instabilities
\citep{boss_1997, boss_2011}. At larger separations, disk
fragmentation at the Class 0/I stages seems to be more likely
\citep{kratter_etal2010}. Due to their large separations from their
host stars, these objects are probably the most promising candidates
to directly observe and characterize circumplanetary disks that may
have formed either from the surrounding cloud or from the massive disk
around the host star.  Indirect evidence for the presence of such
disks around several PMCs has been provided recently by the detection
of emission lines or possible mid-infrared excesses that might be
related to disks and outflows \citep[e.g.][]{seifahrt_etal2007,
  schmidt_etal2008, bowler_etal2014}, and large optical/ultraviolet
excess emission, potentially indicating the presence of shocks
produced during the accretion process \citep{zhou_etal2014}.

One of the most convincing candidates for a disk around a
potential PMC has been recently provided by the third component in the
triple system FW\,Tau. This system belongs to the nearby
($d\sim140$\,pc) and young ($\sim2$\,Myr) Taurus-Auriga star forming
region \citep{goldsmith_etal2008}. The primary in FW\,Tau consists of
a close binary, composed of two M5 stars, with a projected separation
of $11$\,au (75\,mas). The faint tertiary component has been first
reported by \citet{white_ghez2001} and confirmed to be co-moving at a
projected separation of 330\,au by \citet{kraus_etal2014}. The primary
binary in FW\,Tau is not showing evidence for the presence of
accretion \citep{cieza_etal2012} while for the third object
indications for accretion have been found \citep{bowler_etal2014}.
 
Moreover, \citet[][hereafter
\citetalias{kraus_etal2015}]{kraus_etal2015} have recently derived a
dust disk mass of 1--2\,$\Mearth$, based on their ALMA band\,6
continuum observations. While their observations also cover the
$^{12}$CO\,(2--1) line, their observational setup used wide velocity
channels (20\,km\,s$^{-1}$) and the line remained undetected. Despite
these constraints however, the nature of the tertiary companion in the
FW\,Tau system has remained unclear, as the near-IR photometry and
spectroscopy is consistent with both a planetary mass object with
accretion-induced veiling, or a brown dwarf or low-mass star with
spectral types M5--M8 embedded in an edge-on disk
\citep{bowler_etal2014}.

Here we present a clear detection of the $^{12}$CO\,(2--1) line
towards the tertiary companion to FW Tau and construct the spectral
energy distribution (SED) of the system from near-IR to the millimeter
wavelengths. Using radiative transfer and CO-line disk modeling we
explore the two proposed scenarios.

\section{ALMA cycle 1 observations: continuum and gas detection\label{sect:observations}}

We observed FW\,Tau with ALMA in Cycle\,1 on 2013 December 4. We
obtained one epoch observation in band\,6, with the correlator
configured to obtain one baseband centered at 230.52\,GHz which was
aimed at detecting the $^{12}$CO\,(2--1) spectral line, and three
continuum basebands centered at 228.52, 214.52, and 212.52\,GHz. The
total bandwidth for the observations was 7.5\,GHz, with a unique
spectral spacing of 488.28\,kHz in 3\,840 channels for each 1.875\,GHz
baseband.

The observations were carried out using 27 antennas in a compact
configuration, implying we obtained baselines ranging from 15.8\,m to
462.9\,m (12.2 to 356.1\,k$\lambda$). Standard calibration steps were
applied to the data. The calibration sources associated to these
observations were QSO~J042315-012033 for band-pass calibration, and
QSO~J051002+180041 for gain and phase calibrations, and one antenna
was flagged-out because of high system temperature. The observations
consisted of 4 scans, which translates to a total time on source of
3.6\,min for the field. After calibrating the set of raw visibilities,
we applied one iteration of self-calibration in order to correct
low-order phase calibration errors.

Deconvolving the set of visibilities with the CLEAN task implemented
in CASA \citep{mcmullin_etal2007} and using natural weighting we
obtained the final images. The continuum image has an rms of
0.18\,mJy/beam, with an elongated beam size of
$0\farcs75\times1\farcs31$ and a position angle of $-48.6\degr$
(north-east). The mean rms associated to the individual
$^{12}$CO\,(2--1) channels is 21.1\,mJy/beam per 0.635\,km\,s$^{-1}$
bin, after removing the continuum contribution in the visibility
domain.

The continuum image shows a clear detection centered on the location
of the faint companion, while it shows no significant emission from
the binary system, confirming the detection reported in
\citetalias{kraus_etal2015}. The total flux from the third object is
$3.4\pm0.2$\,mJy. This detection is above the $1.78\pm 0.03$\,mJy
reported by \citetalias{kraus_etal2015} and in marginal agreement with
the $4.5\pm1.1$\,mJy at 850\,$\mu$m measurement of
\citet{andrews_williams2005}.

\begin{figure}
\begin{center}
\includegraphics[width=10cm, trim = 0mm 0mm 0mm 0mm, clip]{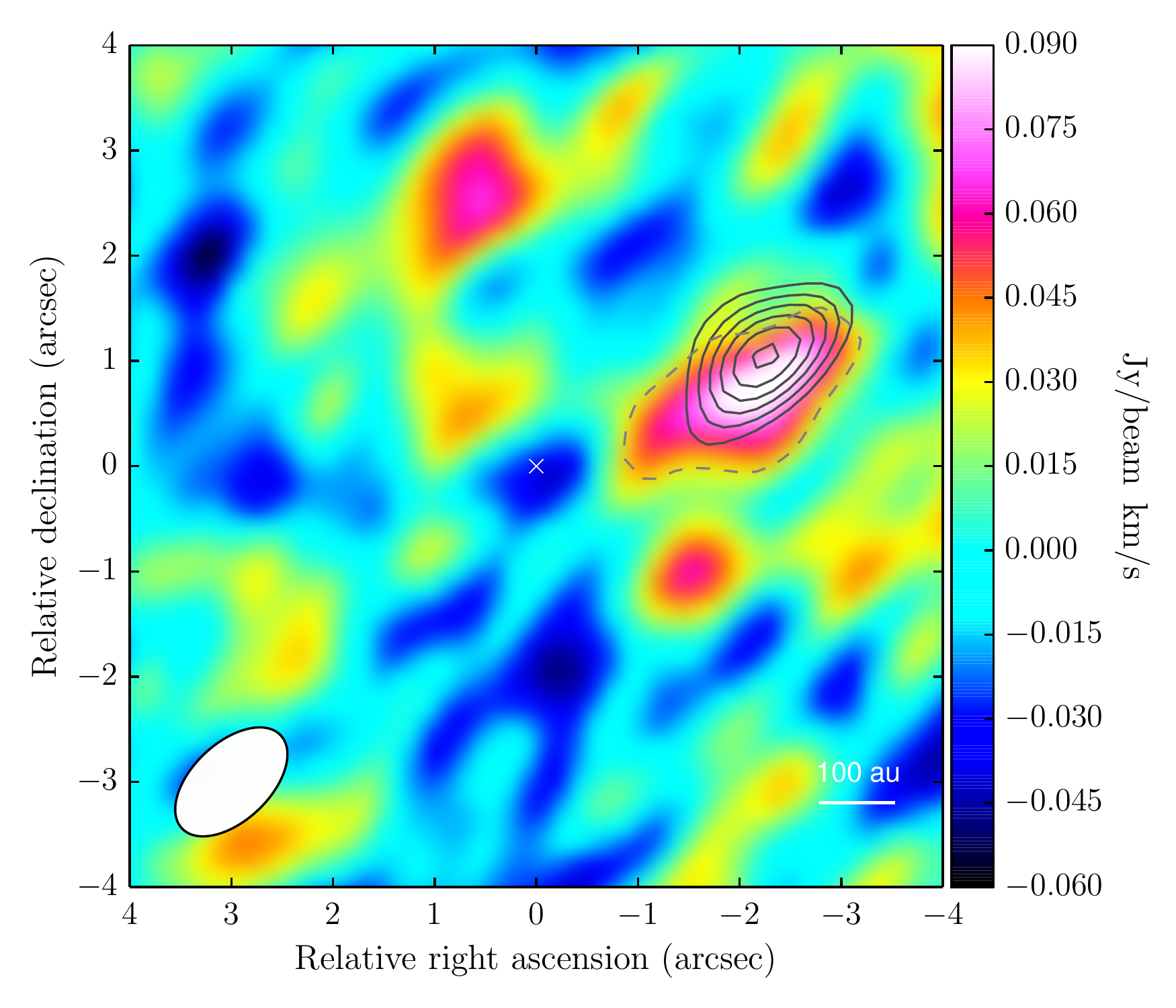}
\end{center}
\caption{ \small%
  Integrated $^{12}$CO(2-1) emission map. The solid-line contours show
  the continuum detection above $5\sigma$ in steps of $2\sigma$. The
  dashed contour limits the $3\sigma$ emission from the
  $^{12}$CO\,(2--1) line used to calculate the line profile. The
    white ellipse shows the synthesized ALMA beam.%
    \label{fig:comap}}
\end{figure}

Similar to the continuum observations, we detected weak emission from
$^{12}$CO\,(2--1) gas located at the same position as the third
object, with a SNR $\sim4$ in three consecutive channels. This
emission appears as a single peak line centered at 5.7\,km\,s$^{-1}$,
with a measured FWHM of 1.9\,km\,s$^{-1}$, and a maximum of 72.8\,mJy,
based on a Gaussian fit to the line profile. The emission peaks from
both the continuum and the integrated spectral line appear separated
by $0.38\pm0.16$\,arcsec (Fig.~\ref{fig:comap}).  Integrating our line
profile over the corresponding channels we obtain a total flux of
156\,mJy\,km\,s$^{-1}$.

\section{Spectral Energy Distributions} 
 
The detection of gas around the third object in FW\,Tau is consistent
with the proposed accreting nature of the disk and provides additional
information to potentially constrain the properties of the third
object.  Before investigating this, we here summarize the
observational data available for the triple system FW\,Tau with
special emphasis on separating resolved and unresolved observations.

The SED of the full FW\,Tau system is dominated by emission from the
photospheres of the central binary at wavelengths shorter than
$\sim$15\,$\mu$m.  However, high resolution observations have also
been able to identify the contribution from the third object in the J,
H, K, and L bands \citep{kraus_etal2014}. Emission at 1.3\,mm is also
confirmed as originating from the third component, both in this paper
and in previous observations \citepalias{kraus_etal2015}. This study
also find a lack of 1.3\,mm flux from the central binary, allowing
upper-limits to be derived for this component.

The excess emission at wavelengths longer than 15\,$\mu$m and shorter
than 1.3\,mm is unresolved, i.e. it could stem from a circumbinary
disk around the central binary and/or a disk around the third
object. These unresolved observations cover the \emph{Spitzer}
photometry from \citet{cieza_etal2012}, the \emph{Herschel} data from
\citet{howard_etal2013}, and the 450\,$\mu$m upper limit and 850
$\mu$m detection from \citet{andrews_williams2005}.  Given the 1.3\,mm
detections of the third object, it is very likely that the later
detection is also associated with it. Table \ref{table:photometry}
summarizes the photometric data available for the triple system
FW\,Tau separating their different contributions.

\section{Disk models}

The first attempts to understand the nature of the third object in
FW\,Tau object rapidly led to the hypothesis of it being a PMC
\citep{kraus_etal2014}.  However, as discussed in detail in
\citetalias{kraus_etal2015} it might also be a late-type stellar or
sub-stellar companion embedded in an edge-on disk. We here confront
both scenarios with the complete set of available observations by
presenting an example model for each case.  A complete parameter study
is beyond the scope of this letter and not warranted given the limited
quality of the current data.

For the SED analysis we assumed an extinction towards the line of
sight of $A_V=0.4$ \citep{kraus_etal2014}. We applied this correction
to both the binary and third object photometry.  The binary
contribution was assumed to be the sum of two identical M5 dwarfs
normalized to the J-band flux, with temperatures of $\Teff=3200$\,K,
masses of $M=0.22\,\Msun$ and a distance of 140\,pc. These assumptions
imply both stars have a radius of 1.1\,$\Rsun$, suggesting an age of
$\sim2$\,Myr \citep{baraffe_etal2015}, which is in agreement with the
Taurus age.

\subsection{An edge-on disk around a substellar
  object\label{sect:edgeon}}
Assuming all the excess emission comes from the third object, the SED
resembles that of an edge on disk around a (sub)stellar body. To model
this possibility, we removed from the observed excesses the
contributions from the central binary emission. Using the radiative
transfer code MCFOST \citep{pinte_etal2006, pinte_etal2009} we find
that for inclinations $i\gtrsim85$\,deg the extinction on the central
object is too high to agree with a spectral type later than M5 as
required from spectroscopy \citep{bowler_etal2014}. Therefore, we
focus on models with inclinations in the range of 70--85 deg, in which
the disk has a direct effect on the central object emission as seen
from our line of sight without causing too much extinction.

We explore various parameters for reproducing the SED, i.e. the disk
dust mass $\mathrm{M}_\mathrm{d}$, the scale height at 100\,au $H_0$,
flaring angle $\beta$, grain size distribution ($a_\mathrm{min}$,
$a_\mathrm{max}$) for silicate composition, and the surface density
exponent $\gamma$ for a power-law disk geometry. We also explore
different values for the temperature and luminosity of the central
object. We fixed the size of the disk to be 100\,au in diameter, in
agreement with the maximum allowed by the non-resolved ALMA
detections. The continuum emission is only weakly dependent on this
assumption as the dust is optically thick only to a few tens of au,
thus a smaller disk will still remain as a plausible alternative with
small effects on the model SED. However, the gas emission is optically
thick throughout, and will be affected by the assumed disk size. For
all high inclination configurations we identify a degeneracy between
the scale height, flaring exponent, inclination, and central object
mass, which makes the modeling of this disk particularly
challenging. Nonetheless, we find that for most parameter combinations
a dust mass of $\sim3\,\Mearth$ and a disk inner radius of
$\sim 1$\,au are required. Similarly, we find that an inclination
close to $i\simeq80.5$\,deg is required, as slightly lower
inclinations expose too much stellar flux and a slightly higher values
extinct too much, making the model inconsistent with the observed
SED. Although the mass of the central object is less constrained by
the SED, it is possible to find reasonable representations of the SED
for masses $\lesssim0.14\,\Msun$.

\begin{figure}
\begin{center}
\includegraphics[width=0.8\textwidth, trim = 0mm 0mm 0mm 0mm, clip]{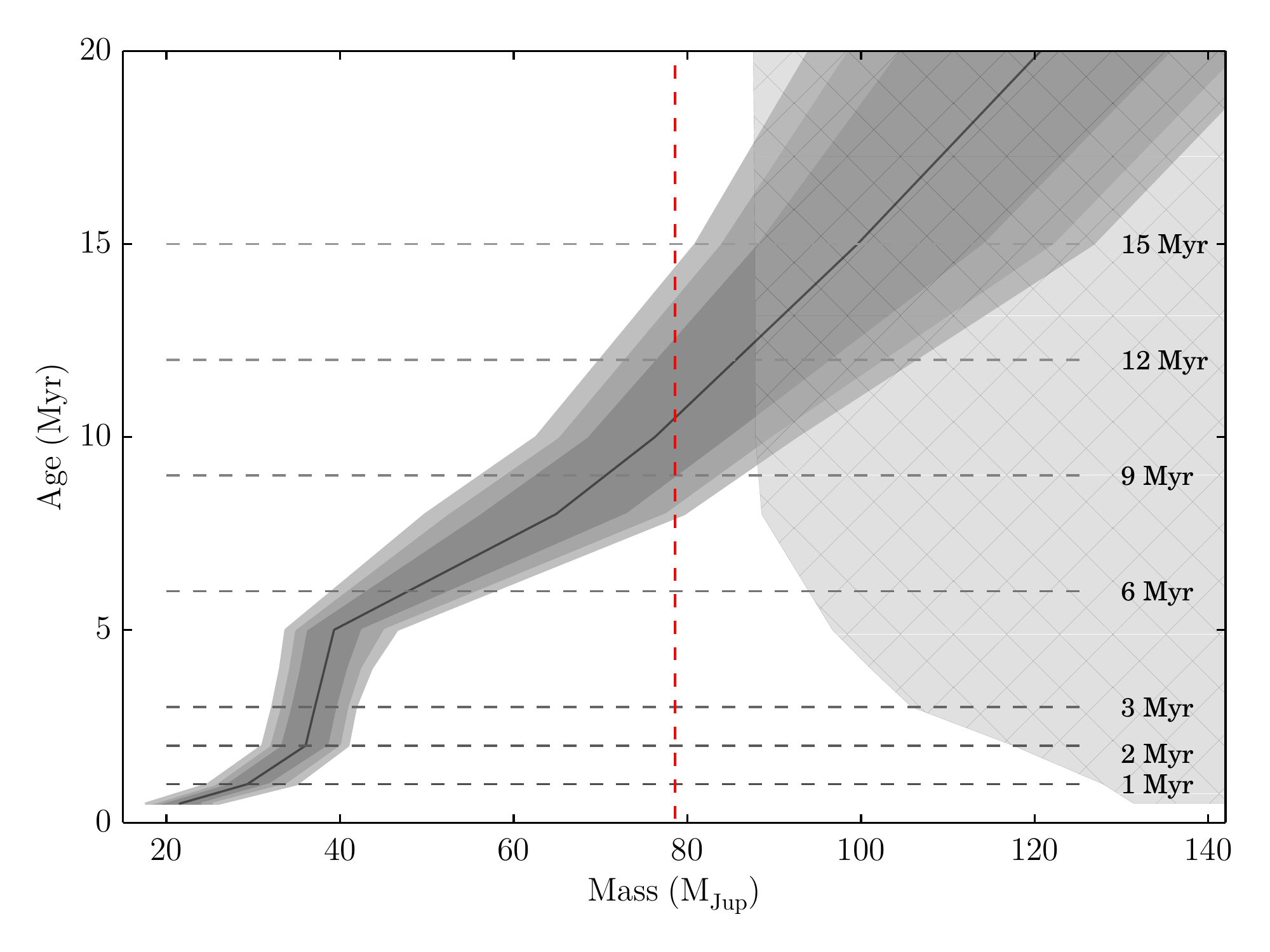}
\end{center}
\caption{ \small%
  Age-mass region constrained by the edge-on modeling. Gray
    regions represent the $1-3\sigma$ range of allowed ages for a given central
    object mass. The red-dashed line indicates the deuterium-burning
    limit mass for a solar composition \citep{chabrier_etal2000}. The
    shaded area limits the maximum allowed mass given spectroscopic
    constraints \citep{bowler_etal2014}.%
    \label{fig:mass_age}}
\end{figure}

However, we can constrain the mass of the central object in the
edge-on model if we require its age to be close to the one estimated
for the central binary. The larger the assumed mass for the central
object, the older it must be to reach agreement with the
observations. Figure\,\ref{fig:mass_age} shows this dependence,
assuming the evolutionary models of \citet{baraffe_etal2015} and the
calculated photospheric luminosity of
$9\times10^{-3}$\,L$_\odot$. Even a conservative upper-limit of 6\,Myr
($\sim3$ times the estimated age of the central binary) would imply
the mass of the third object to be $\lesssim60\,\Mjup$. We can
therefore rule out stellar masses but the third object might be a
forming brown dwarf seen edge-on.

The detected $^{12}$CO\,(2--1) emission might provide a further
observational constraint on the nature of the third body. Therefore,
we created an MCFOST radiative transfer model for the detected gas
emission, assuming an ISM CO/H$_2$ abundance, a freezing temperature
of 20\,K \citep{qi_etal2004, degregoriomonsalvo_etal2013}, turbulent
velocities ranging from 0.1--0.2\,km\,s$^{-1}$, and the same spectral
resolution as provided by the observations. We found that, given the
low signal to noise of the line and the effects of photodissociation
by the third object and the central binary \citep{williams_best2014},
the total gas content of the disk remains an observationally
unconstrained parameter.

Moreover, the shape of the line does not provide unambiguous
observational constraints on the mass of the central object. For the
spectral resolution given by our observations, masses
$\gtrsim35\,\Mjup$ predict double-peaked lines in disagreement with
the observed single-peaked line. However, we can not exclude such
masses as the high noise level of the detection could have smeared out
the line splitting.  In addition, the observed line emission could be
significantly affected by cloud contamination. Indeed, given that the
measured velocity of the third object in FW\,Tau is close to the mean
velocity of the Taurus cloud, i.e. $\sim7$\,km\,s$^{-1}$
\citep{goldsmith_etal2008}, it is possible that either the blueshifted
or redshifted emission from the disk and the cloud appear fused in the
interferometer, thus precluding the interferometer from resolving the
individual contributions and filtering them out \citep[see
e.g.][]{canovas_etal2015}.  The same effect could explain the small
(2.4$\sigma$) difference in position of the centers of the continuum
and line detections (see \S\ref{sect:observations}).

As an example for an edge-on model that explains all the available
observations, we show a model fit to the SED and the CO line assuming
a 5\,Myr old $40\,\Mjup$ third object in Fig.\,\ref{fig:models}, where
it is clear that all the observed excess flux can be explained by
material confined to the third object.

\subsection{A disk around a planetary mass companion
  object\label{sect:faceon}}

The third object in FW\,Tau could also be a cooler low-mass object
with a disk seen at low inclination.  Indeed, the optical and near-IR
emission is well fitted by a $\Teff\sim1800$\,K substellar model, with
a luminosity implying a 1--2\,Myr old age, in agreement with the age
of Taurus forming objects, and consistent with a 5--7\,$\Mjup$ mass
body \citep{baraffe_etal2015}.

The projected separation of the third object from the binary has been
measured to be $330$\,au, and if the true separation is close to the
projected one, the flux from the nearby binary will contribute to disk
heating, thus affecting the temperature structure and consequently its
emission. To reproduce the observed SED we have modeled the disk
around the third component assuming the binary contribution to be
identical to that described in \S\ref{sect:edgeon}. The separation
between the binary stars was assumed to be 11\,au, whereas the
distance and the relative orientation of the binary with respect to
the disk were changed in order to explore different configurations for
the external illumination.  Apart from this additional parameter we
explored the same parameter space as in \S\ref{sect:edgeon}.

\begin{figure}
\includegraphics[width=\textwidth, trim = 0mm 0mm 0mm 0mm, clip]{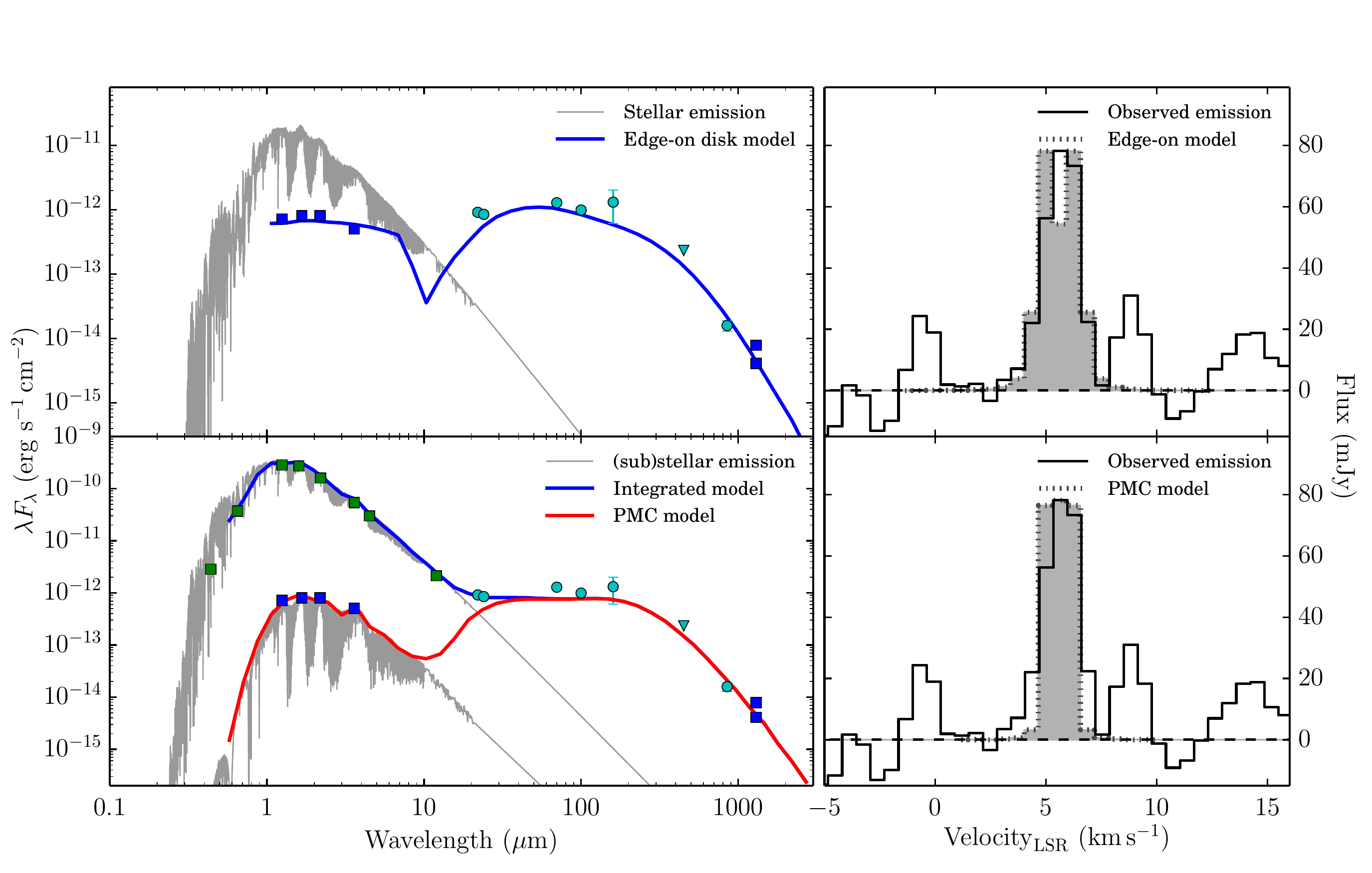}
\caption{\small%
  {\sl Left panels}: SED for the FW\,Tau system. Green squares
  represent the emission from the central binary, blue squares
  represent the emission associated to the third component, cyan
  circles and triangle are the unresolved photometric observations and
  upper limit respectively. Upper-left panel shows the edge-on model
  described in \S\ref{sect:edgeon} (blue solid line), including the
  highly absorbed substellar photosphere (gray solid line). Lower-left
  panel shows the PMC model (red solid line) including the
  photospheric emission from a 1800\,K central object, and the
  integrated SED model composed of the PMC model plus the photospheric
  emission from the central binary, shown as M5-type photospheres
  (\S\ref{sect:faceon}). The parameters used to create these models
  are specified in Table~\ref{table:models}. %
  {\sl Right panels}: Detected ALMA $^{12}$CO\,(2--1) line profile
  measured over the $3\sigma$ region shown in Fig.~\ref{fig:comap}
  (black solid line). The dotted lines represent the edge-on model
  (\S\ref{sect:edgeon}; upper-right panel), and the planetary mass
  companion model (\S\ref{sect:faceon}; lower-right panel),
  respectively. The unconstrained total gas mass was adjusted to fit
  the maximum detected emission.%
  \label{fig:models}}
\end{figure}

We find that the SED of FW\,Tau can be explained with models in which
the binary illuminates the disk with an incident angle
$\sim 20-50$\,deg with respect to the disk plane as these orientations
maximize the heating of the disk (Fig.~\ref{fig:models}). Other
configurations do not expose the disk to enough radiation, creating a
lack of emission at longer wavelengths. We find that
$\sim2-3\,\Mearth$ of dust are enough to reproduce the emission at
(sub) mm-wavelengths, in agreement with the estimates of
\citetalias{kraus_etal2015}. The mid/far-IR emission is consistent with
very small flaring angle exponents (i.e. $\beta\simeq1.0$), and an
inner radius of $\sim0.3$\,au.  The overall flux excess is only
achieved, however, by introducing a relatively large scale height
($\sim 22-25$\,au at 100\,au), and models with smaller values tend to
underestimate the mid/far-IR emission.

Concerning the gas line, for an assumed central body mass of
7\,$\Mjup$, the spectral profile appears as a clear single line for
inclinations $i\lesssim15$\,deg, in agreement with the observations
(see Fig.\,\ref{fig:models}), and the Keplerian double-peak appears
only for larger inclinations. The amplitude of the CO line peak
suggests a very small gas mass, because the amplitude of the line
predicted by the model, for an ISM CO/H$_2$ abundance and gas-to-dust
mass ratio, is much larger than the observed values. This could be
either an effect of CO photodissociation which can reduce the CO/H$_2$
abundance \citep{visser_etal2009,williams_best2014}, an intrinsic very
low gas-to-dust mass ratio, and/or a smaller disk size because a
smaller emitting area would predict less CO flux and thus require a
less reduced gas content.  Similar to the high-inclination case, it is
also possible that the lack of flux from the observations compared to
the model might be due to the interferometer filtering out a major
fraction of the flux.

\section{Conclusion}

We present ALMA Cycle~1 band~6 continuum and $^{12}$CO\,(2--1)
observations of the triple system FW\,Tau and test two different
scenarios against the new observations.

We find that the edge-on model can recreate all observed excess
emission. Although the observed single-peaked line is most naturally
reproduced assuming a low mass for the third object, larger masses can
not be excluded because of the low SNR of the detection and the
possibility of cloud contamination.  Assuming the third object and its
central binary are coeval however, evolutionary tracks strongly
suggest it is a substellar object.

The low inclination scenario for a planetary mass object, in
combination with the radiation received from the close binary, is
capable of reproducing the SED and the single-peaked line
emission. The weakness of the $^{12}$CO\,(2--1) line either indicates
a low gas mass, significant photodissociation affecting the CO/H$_2$
abundance, significant contamination by the cloud, a smaller disk, or
a combination of these four effects are at work.

Resolved continuum observations of both the binary and third
components of the FW\,Tau system, as well as resolved optically thin
line detections including both CO higher-transition and isotopologues,
will definitively unveil the nature of this intriguing
system. Regardless, all data suggest FW\,Tau is a substellar object
caught in formation, cementing its position as a vital object for
understanding the brown dwarf/planet formation process.

\acknowledgments{%
  C.C. and M.R.S. acknowledge the support from CONICYT FONDECYT
  grants 3140592 and 1141269 respectively. C.C., and H.C. acknowledge
  support from ALMA-CONICYT grant 31130027. C.C., A.H., M.R.S., H.C.,
  L.C. acknowledge support from the Millennium Nucleus RC130007
  (Chilean Ministry of Economy). This paper makes use of the following
  ALMA data: ADS/JAO.ALMA\#2012.1.00350.S. ALMA is a partnership of
  ESO (representing its member states), NSF (USA) and NINS (Japan),
  together with NRC (Canada), NSC and ASIAA (Taiwan), and KASI
  (Republic of Korea), in cooperation with the Republic of Chile. The
  Joint ALMA Observatory is operated by ESO, AUI/NRAO and NAOJ. The
  National Radio Astronomy Observatory is a facility of the National
  Science Foundation operated under cooperative agreement by
  Associated Universities Inc. This publication makes use of data
  products from the Wide-field Infrared Survey Explorer, which is a
  joint project of the University of California, Los Angeles, and the
  Jet Propulsion Laboratory/California Institute of Technology, funded
  by the National Aeronautics and Space Administration.%
}


\newpage

\begin{deluxetable}{cccccc}
\tablewidth{0pt}
\footnotesize%
\tablecaption{FW\,Tau photometry data\label{table:photometry}}%
\tablehead{\colhead{Wavelength}&\colhead{Flux}&\colhead{Flux}& \colhead{Flux}&\colhead{Unit}&\colhead{Reference$^a$}\\
            \colhead{($\mu$m)}&\colhead{FW Tau A/B}&\colhead{Third component}&\colhead{Unresolved} &\colhead{} 
}
\startdata  
   1.25   &      10.34      &   17.34\,$\pm$\,0.07  &      \nodata        &  mag  &    1      \\
   1.66   &       9.67      &   16.24\,$\pm$\,0.07  &      \nodata        &  mag  &    1      \\
   2.20   &       9.38      &   15.29\,$\pm$\,0.07  &      \nodata        &  mag  &    1      \\
   3.35   &     \nodata     &   \nodata             & 9.196\,$\pm$\,0.026 &  mag  &    2      \\ 
   3.6    &     \nodata     &   \nodata             &  64.8\,$\pm$\,0.3   &  mJy  &    3      \\ 
   3.8    &       9.19      &   14.25\,$\pm$\,0.10  &      \nodata        &  mag  &    1      \\ 
   4.5    &     \nodata     &   \nodata             &  44.8\,$\pm$\,0.1   &  mJy  &    3      \\ 
   4.6    &     \nodata     &   \nodata             &  41.5\,$\pm$\,1.0   &  mJy  &    2      \\ 
   5.8    &     \nodata     &   \nodata             &  32.4\,$\pm$\,0.2   &  mJy  &    3      \\ 
   8.0    &     \nodata     &   \nodata             &  18.0\,$\pm$\,0.1   &  mJy  &    3      \\ 
  12      &     \nodata     &   \nodata		    &   8.2\,$\pm$\,0.3   &  mJy  &    2      \\ 
  22      &     \nodata     &   \nodata             &     $<10.5$         &  mJy  &    2      \\ 
  24      &     \nodata     &   \nodata             &  6.79\,$\pm$\,0.4   &  mJy  &    4      \\ 
  70      &     \nodata     &   \nodata             &    30\,$\pm$\,4     &  mJy  &    5      \\
 100      &     \nodata     &   \nodata             &    33\,$\pm$\,4     &  mJy  &    5      \\
 160      &     \nodata     &   \nodata             &    70\,$\pm$\,40    &  mJy  &    5      \\
 450      &     \nodata     &   \nodata             &     $<30.0$         &  mJy  &    6      \\
 850      &     \nodata     &   \nodata             &   4.5\,$\pm$\,1.1   &  mJy  &    6      \\
1300      &    $<0.084$     &  1.78\,$\pm$\,0.03    &      \nodata        &  mJy  &    7      \\
1300      &     \nodata     &  3.4\,$\pm$\,0.2      &      \nodata        &  mJy  & This work \\
\enddata
\tablecomments{ $^a$References: (1) \citet{kraus_etal2014}; (2)
  \citet{wright_etal2010}; (3) \citet{cieza_etal2009}; (4)
  \citet{cieza_etal2012}; (5) \citet{howard_etal2013}; (6)
  \citet{andrews_williams2005}; (7) \citet{kraus_etal2015}.  }
\end{deluxetable}

\newpage

\begin{deluxetable}{cccc}
\tablewidth{0pt}
\footnotesize%
\tablecaption{Parameters in the models\label{table:models}}%
\tablehead{\colhead{Parameter}&\colhead{PMC model} & \colhead{Edge-on disk model} & \colhead{Unit}}
\startdata
$i$                    & $10$        & $80.5$        & deg\\
$M_\mathrm{S}$           & 7         & 40          & $\Mjup$\\
$M_\mathrm{disk}$        & 3         & 3           & $\Mearth$\\
$H_0$                   & 25        & 15          & au\\
$\beta$                 & 1.00      & 1.25        & \nodata \\
$\gamma$                & -0.65     & -0.1        & \nodata \\
$a_\mathrm{min}$          & $1.0\times10^{-2}$ & $1.0\times10^{-2}$ & $\mu$m\\
$a_\mathrm{max}$          & $1.0\times10^{3}$  & $1.0\times10^{3}$  & $\mu$m\\
\enddata
\tablecomments{These are the parameters used to create the models shown in Fig.~\ref{fig:models}.}
\end{deluxetable}

\newpage

\newpage

\newpage

\end{document}